\documentclass[aps,twocolumn,floatfix,footinbib,groupedaddress,superscriptaddress,showpacs,notitlepage]{revtex4-1}
\usepackage[dvips]{graphicx}
\usepackage{amssymb,amsfonts,amsmath}
\usepackage{mathrsfs}

\newcommand{\bra}[1] {\left\langle #1 \right|}
\newcommand{\ket}[1] {\left| #1 \right\rangle}
\def\openone{\leavevmode\hbox{\small1\kern-4.2pt\normalsize1}}

\newcommand{\bracket}[2]{\langle#1|#2\rangle}
\newcommand{\nn}{\nonumber\\}
\newcommand{\bea}{\begin{eqnarray}}
\newcommand{\ea}{\end{eqnarray}}

\begin{document}
\title{Quantum simulator of an open quantum system using superconducting qubits: 
exciton transport in photosynthetic complexes }

\author{
Sarah Mostame$^1$, 
Patrick Rebentrost$^1$, 
Alexander Eisfeld$^1$,
Andrew J. Kerman$^2$
Dimitris I. Tsomokos$^3$, 
and Al\'an Aspuru-Guzik$^*\, $
}

\address{
Department of Chemistry and Chemical Biology, Harvard University, Cambridge, Massachusetts 02138, USA
\\
$^2$Lincoln Laboratory, Massachusetts Institute of Technology, Lexington, Massachusetts, 02420, USA
\\
$^3$Department of Mathematics, Royal Holloway, University of London, Egham, TW20 0EX, United Kingdom
}

\begin{abstract}
Open quantum system approaches are widely used in the description of
physical, chemical and biological systems.
A famous example is electronic excitation transfer in
the initial stage of photosynthesis, where harvested energy
is transferred with remarkably high efficiency to a reaction center.
This transport is affected by the motion of a structured vibrational environment,
which makes simulations on a classical computer very demanding.
Here we propose an analog quantum simulator of complex open system dynamics
with a precisely engineered quantum environment.
Our setup is based on superconducting circuits, a well established technology.
As an example, we demonstrate that it is feasible  to simulate exciton transport in the 
Fenna-Matthews-Olson photosynthetic complex.
Our approach allows for a controllable single-molecule simulation and the investigation of 
energy transfer pathways as well as non-Markovian noise-correlation effects.
\end{abstract}

\maketitle

%%%%%%%%%%%%%%%%%%%%%%%%%%%%%%%%%%%%%%%%%%%%%%%%%%%%%%%%%%%%%%%%%%%%%%%%%

% \date{\today}
% 
%%%%%%%%%%%%%%%%%%%%%%%%%%%%%%%%%%%%%%%%%%%%%%%%%%%%%%%%%%%%%%%%%%%%%%%%%%
%%                      Introduction
%%%%%%%%%%%%%%%%%%%%%%%%%%%%%%%%%%%%%%%%%%%%%%%%%%%%%%%%%%%%%%%%%%%%%%%%%%
%----------------------------------------------------------
Understanding strongly interacting quantum systems with many degrees of 
freedom is one of the big challenges in physics and chemistry \cite{ivan2011}.
Classical computational methods are restricted by exponentially increasing 
amount of resources required for the simulations.
Quantum computers are conjectured to be a possible solution as the resources to simulate 
arbitrary quantum systems grow polynomially with the size of the system under study
\cite{feynman,nielsen2000}.
However, universal quantum computers of sufficient size and performance are not available 
yet, one of the big problems being the loss of quantum mechanical coherence, i.e., 
decoherence \cite{divincenzo}.
Designing a special quantum system in the laboratory, which mimics the quantum 
dynamics of a particular model of interest, see for example, 
Refs.~\cite{nori,alan2005,ralf2005,sarah2008,tsomokos2010,geller},  can be a more viable 
alternative to an all-purpose quantum computer.

Here we propose a quantum simulator architecture using superconducting quantum bits 
(qubits) that is capable of simulating complex open quantum systems using currently-available 
technology in realizable parameter ranges.
We will focus on the Fenna-Matthews-Olson (FMO) 
pigment-protein complex on a single molecule level. 
The recent observations \cite{engel2007,lee2007}
of quantum beatings and long-lived quantum coherence in several
photosynthetic light-harvesting complexes, such as the  FMO
complex in the green sulfur bacterium {\em Chlorobium tepidum} or the reaction 
center of the purple bacterium {\em Rhodobacter sphaeroides}, suggest
possible evidence that quantum effects give rise to the high energy
transport efficiency found for these complexes. 
There is a remarkable amount of recent theoretical research related to the 
question of the molecular structure, vibrational environment, 
the origin and the role of long-lived quantum coherences
\cite{cho2005,adolphs2006,muh2007,
patrick2009NJP,patrick2009JPC,plenio2009,mukamel2009,ishizaki2009,
ishizaki2010,patrick2011JCP,renger2011}.
The electronic degrees of freedom are coupled to a finite temperature vibrational environment 
and the dynamics of the relevant electronic system can be studied by means of open quantum system approaches.
In quantum computing, the focus of much of the research has been on 
reducing the magnitude and influence of environmental decoherence and dissipation.
However, controlled coupling to a dissipative environment can also be exploited \cite{verstraete2009,barreiro2011,yu2010}.
In this work, we focus on {\em engineering} the decoherence to simulate open quantum 
systems that are challenging to study using classical computers. 

We propose two approaches for simulating the vibrational environment. 
The first approach is based on engineering a classical noise source 
such that it represents the atomistic fluctuations of the protein environment. 
A prototypical experiment of environment-assisted 
quantum transport (ENAQT) can be performed \cite{patrick2009NJP}.
The second approach allows for the precise engineering of the 
complex non-Markovian environment, i.e., 
an environment that has long-term memory. 
This is achieved by the explicit coupling of quantum inductor-resistor-capacitor~(LRC) 
oscillators to the qubits which allows for energy and coherence exchange between the 
resonators and the qubits.
Both approaches are based on present-day superconducting qubit implementations.
Fabrication of superconducting circuitry is done by several research labs.
We focus here on flux qubits, where two-qubit coupling was shown to be sign- and magnitude-tunable 
\cite{lloyd2007,paauw2009} and methods of scaling to a moderate number of qubits have been discussed 
in Refs \cite{hime2006,wilhelm2007}.
We show that realistic simulation of photosynthetic energy transfer is feasible with current 
superconducting circuit devices. 
%

%%%%%%%%%%%%%%%%%%%%%%%%%%%%%%%%%%%%%%%%%%%%%%%%%%%%%%%%%%%%%%%%%%%%%%%%%%
%%                      The model Hamiltonian / FMO
%%%%%%%%%%%%%%%%%%%%%%%%%%%%%%%%%%%%%%%%%%%%%%%%%%%%%%%%%%%%%%%%%%%%%%%%%%
%
\section{The model Hamiltonian}

%In this section we introduce the general open quantum system Hamiltonian to be  
%simulated and specialize it to the FMO complex.
%
We are interested in the dynamics of a finite dimensional system which is linearly 
coupled to a bath of harmonic oscillators.
In the following we refer to the system as ``electronic system''
and to the quantum environment as ``phonon bath'' or ``vibrational environment''.
The corresponding total Hamiltonian is written~as
\bea
\label{tothamilton}
H_{\mathrm{tot}}= H_{\mathrm{el}}+H_{\mathrm{ph}}+H_{\mathrm{el-ph}}\, .
\ea
%
%%%%%%%%%%%%%%%%%%%%%%%%%%%%%%%%%%%%%%%%%%%%%%%%%%%%%%%%%%%%%%%%%%%%%%%%%%

\subsection{The system}
We are often (e.g., in the FMO complex) interested in the transfer of a single 
electronic excitation.
Thus basis states $\ket{j} $ are defined by the electronic excitation residing on molecule (site) $j$ and 
all other sites being in their electronic ground state.
The electronic Hamiltonian in this site basis is given by
\mbox{
$ H_{\mathrm{el}}=\sum_{j=1}^{N}\tilde{\varepsilon}_j \ket{j}\bra{j}
+\sum_{i<j}^{N}V_{ij}
\left(\ket{i}\bra{j}+\ket{j}\bra{i} \right) \,,$}
\cite{may-kuehn}.
The diagonal energies $\tilde{\varepsilon}_j $ are identified with the 
electronic transition energies of site $j$ and the off-diagonal elements $V_{ij}$ 
are the intermolecular (transition-dipole-dipole) couplings between sites $i$ and $j$.
Different local electrostatic fields of the protein at different sites shift the electronic transition 
energies \cite{muh2007}, resulting in a complicated energy landscape.
%

%%%%%%%%%%%%%%%%%%%%%%%%%%%%%%%%%%%%%%%%%%%%%%%%%%%%%%%%%%%%%%%%%%%%%%%%%%
\subsection{Coupling to the quantum environment}
\begin{figure}

\centering
\resizebox{0.98\linewidth}{!}{\includegraphics{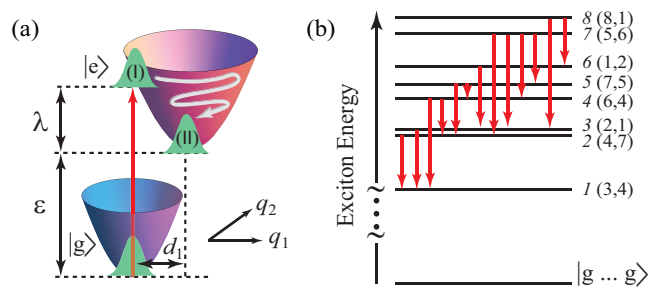}}
\caption{
(a) Model of exciton-phonon dynamics in a system with two electronic
states $\ket{g}$ and $\ket{e}$ and two dissipative vibrational modes $q_1$ and
$q_2$. 
After a vertical Franck-Condon electronic excitation the phonon 
degrees of freedom are in a non-equilibrium position (I) from which the system
relaxes to the displaced equilibrium configuration in the excited
state (II). 
The displacements are given by $d_1$ and $d_2$ and the energy scale associated with this 
relaxation is the reorganization energy $\lambda$. 
This vibrational reorganization is not captured in most Markovian models. 
(b) Energy levels of the electronic Hamiltonian for the FMO complex denoted
by $M (i,j)$, where $(i,j)$ indicate the two most significant BChl pigments participating
 in the delocalized excitonic states $M$.
The red arrows indicate the dominant pathways for the energy transport based on Redfield 
theory \cite{patrick2009JPC}.
}
\label{fig:reorg}
\end{figure}

The vibrational environment is represented by a set of displaced harmonic oscillators.
The Hamiltonian of the phonon bath is written as
$H_{\mathrm{ph}}=\sum_{j=1}^{N}H_{\mathrm{ph}}^{j}\, ,$
where \mbox{$H_{\mathrm{ph}}^{j}= \sum_\ell \hbar \omega_\ell ^j ( {a_\ell^j}^\dagger {a_\ell^j} +1/2)$}
with ${a_\ell^j}^\dagger$ (${a_\ell^j}$) 
being the creation (annihilation) operator of excitations in the  $\ell$-th bath mode of site $j$.
In the present work we restrict to the situation where each site has its 
own phonon environment which is uncorrelated with the phonon modes at the other
sities.
This is motivated by recent results obtained for the FMO complex 
\cite{olbrich2011,shim2012}.
The diagonal part of the electronic Hamiltonian couples linearly to the phonon modes. 
 The electron-phonon coupling term can be written as
\bea
H_{\mathrm{el-ph}}=\sum_{j=1}^{N}H_{\mathrm{el-ph}}^{j}=
\sum_{j=1}^{N} \ket{j}  \bra{j} \left[\sum_\ell \, \chi_{j\ell}\,
({a_\ell^j}^\dagger + {a_\ell^j} ) \right] \, .
\nonumber \\
\ea
Here  \mbox{$\chi_{j\ell}=\hbar \omega_\ell^j \, d_{j\ell}$} is the coupling between the $j$-th site
and the $\ell$-th phonon mode with $\omega _{\ell }^{j}$ being the frequency of the $\ell $-th 
phonon mode coupled to the $j$-th site and $d_{j\ell}$ is the dimensionless 
displacement of the minima of the ground and excited 
state potentials  of the $\ell$-th phonon mode at site $j$.
Notice that the so-called reorganization energy 
\mbox{$\lambda_{j}\equiv\sum_\ell \hbar \omega_\ell^j \, d_{j\ell}^2/2$} was implicitly included in the 
above electronic transition energy \mbox{$\tilde{\varepsilon}_j = \varepsilon_j+  \lambda_{j}$}, 
with $\varepsilon_j$ being the energy difference of the minima of the potential 
energy surfaces for site $j$,  see Figure~\ref{fig:reorg}~(a) and the Supporting Information for more details.
%
%%%%%%%%%%%%%%%%%%%%%%%%%%%%%%%%%%%%%%%%%%%%%%%%%%%%%%%%%%%%%%%%%%%%%%%%%%
%------------------------   F I G  1
%%%%%%%%%%%%%%%%%%%%%%%%%%%%%%%%%%%%%%%%%%%%%%%%%%%%%%%%%%%%%%%%%%%%%%%%%%

It is known that complete information about the effect of the environment
on a quantum system is determined by the spectral density (SD) function
\cite{leggett1987}, which is defined by 
\bea
\label{spectral_density}
J_j(\omega)=\sum_\ell |\chi_{j\ell}|^2\delta(\omega-\omega_{j\ell})\,
\ea
 for site~$j$.
 Due to the high number of modes of the environment, 
$J_j(\omega)$ can be considered as a continuous function of $\omega$.
To account for finite temperature, we transform the spectral density \cite{egorov1999,eisfeld2011}  
$C_j(\omega,T)=\left\{1+\coth\left[\hbar\omega/(2 k_BT) \right]\right\} J_j^{\rm A}(\omega)$,
where the subscript ``A'' denotes the antisymmetric spectral density
$J_j^{\rm A}(\omega)= J_j(\omega)$  if $\omega \ge 0\, $; and
$J_j^{\rm A}(\omega)= - J_j(-\omega)$ if $\omega < 0\,$.
The spectral density $C_j(\omega,T)$ fulfills the detailed balance condition 
\cite{egorov1999} and we name it ``temperature-dependent spectral density''.

It turns out that the relevant spectral densities of our problem can be approximated
by a finite number of broadened peaks.
These broadened peaks can often be associated with the vibrational modes of the molecules.
Upon electronic excitation of a molecule, the vibronic Gaussian wavepacket of the ground state 
is projected into a displaced wavepacket in the excited state at the  
Franck-Condon point,  see Figure \ref{fig:reorg}~(a).
The nuclei wavepacket then moves on the excited state potential energy surface and 
reorganizes to the minimum point while the reorganization energies of the respective modes are dissipated.

Finally, to facilitate the comparison with flux qubits we rewrite
the total Hamiltonian \ref{tothamilton} using Pauli matrices 
\bea
\label{total_hamiltonian-02}
H_{\rm{tot}} &=&
\frac{1}{2} \sum_{j=1}^N \, \tilde{\varepsilon}_{j} \, \sigma_z^j 
+ \frac{1}{2} \sum_{i<j}^{N} V_{ij} (\sigma_x^i\sigma_x^j + \sigma_y^i\sigma_y^j)
\nn
&&
+ \sum_{j=1}^N \sum_\ell \hbar \omega_\ell ^j 
\left( {a_\ell^j}^\dagger {a_\ell^j} +\frac{1}{2}\right)
\nn
&&
+ \sum_{j=1}^N \sum_\ell \chi_{j\ell}\, \sigma_z^j 
\left({a_\ell^j}^\dagger + {a_\ell^j} \right) 
\, .
\ea
Expressing the above Hamiltonian in the system energy eigenbasis,
defined by  \mbox{$H_{\rm el} \ket{M} = E_M \ket{M}$}, we have 
$H_{\rm{tot}} = \sum_M E_M \ket{M}\bra{M} 
+\sum_{M,N,\ell}\mathscr{K}_{MN}^{\ell} \ket{M}\bra{N}({a_\ell^j}^\dagger + {a_\ell^j} )+H_{\mathrm{ph}}$
with $\mathscr{K}_{MN}^{\ell} = \sum_j \bracket{M}{j}\bracket{j}{N} \,\chi_{j\ell}$.
This shows that the system-bath coupling is off-diagonal in the eigenbasis.

%%%%%%%%%%%%%%%%%%%%%%%%%%%%%%%%%%%%%%%%%%%%%%%%%%%%%%%%%%%%%%%%%%%%%%%%%%
\subsection{The classical noise approximation}
Although the main goal of the present paper is to simulate the fully quantum mechanical
Hamiltonian \ref{total_hamiltonian-02}, it is also useful to consider the much simpler (but important) case 
where the quantum environment is replaced by  time-dependent fluctuations of the
transition energies. 
This is the basis of the often employed Haken-Strobl-Reineker (HSR) model for excitation transfer
\cite{HSR}.
Furthermore, atomistic MD/QM/MM simulations \cite{olbrich2011,shim2012} can readily provide noise time-series.
In the classical noise approach, the system dynamics is obtained by averaging over many
trajectories with the time-dependent Hamiltonian
\bea\label{classicHamil}
\tilde{H}_{\rm{tot}}=
\frac{1}{2} \sum_{j=1}^N \,  
\left[\tilde{\varepsilon}_{j}+\delta\tilde{\varepsilon}_{j}(t) \right] \, \sigma_z^j 
+ \frac{1}{2} \sum_{i<j}^{N} V_{ij} (\sigma_x^i\sigma_x^j + \sigma_y^i\sigma_y^j)
\, , \nonumber
\\
\ea
where the influence of the environment is solely contained in the time-dependent 
site energy fluctuations $\delta\tilde{\varepsilon}_{j}(t)$. 
Often, as in the HSR model, it is assumed that the fluctuations are 
uncorrelated Gaussian white noise.
%

%%%%%%%%%%%%%%%%%%%%%%%%%%%%%%%%%%%%%%%%%%%%%%%%%%%%%%%%%%%%%%%%%%%%%%%%%%
%%                     FMO complex
%%%%%%%%%%%%%%%%%%%%%%%%%%%%%%%%%%%%%%%%%%%%%%%%%%%%%%%%%%%%%%%%%%%%%%%%%%
\section{The FMO complex}

The model Hamiltonian \ref{tothamilton} can be used to describe a 
single excitation in the FMO complex.
The FMO complex acts as a highly efficient excitation wire, transferring the energy
harvested by the photosynthetic antennae to a reaction center. 
The FMO trimer has a trimeric structure exhibiting C$_{3}$-symmetry
and each of the monomers consists of a network of eight   \cite{renger2011} 
\mbox{bacteriochlorophyll~a} (BChl~a)  pigment molecules.
Since the coupling between monomers is very small and can be neglected on the 
time-scales of interest, we focus on a single monomer in the following.
The BChl pigments in the monomer are surrounded by a protein environment.
Conformational motions of this protein environment (static disorder) are slow compared to the 
timescale of interest and affect energy levels of the pigments by electrostatic interaction \cite{muh2007}.
The ranges of site energy differences $|\tilde{\varepsilon}_i-\tilde{\varepsilon}_j|$ and couplings 
$V_{ij}$ are given in table \ref{Exp-Para}. 
These parameters lead to the energy spectrum of the FMO monomer given in Figure~ \ref{fig:reorg}~(b).

In the present work, we consider two spectral densities relevant to
the FMO complex. 
First, a model super-Ohmic SD \cite{adolphs2006}, 
\mbox{$J(\omega)=\lambda \left(\omega/ \omega_c\right)^2 \,
\exp(-\omega/\omega_c)$}  with reorganization energy 
\mbox{$\lambda=35\,\, \mathrm{cm}^{-1}$} 
and cutoff frequency \mbox{$\omega _{c}=150\,\,\mathrm{cm}^{-1}$},
shown as the transformed $C(\omega,T)$ by the blue dashed line in Figure~\ref{fig:NMS300}~(a).
We have dropped the subscript $j$ under the assumption that
all 8 sites have same spectral density and reorganization energy.
Second, the experimental spectral density \cite{adolphs2006,wendling2000} shown by blue dashed line in 
Figure~\ref{fig:NMS300}~(b). 
Notice that it is very challenging to simulate the experimental  SD
with current computational methods because of the apparent peaks,
see Figure~\ref{fig:NMS300}~(b), and the mixing of vibrational dynamics caused by the electronic interaction between the sites.
This structured spectral density with strong peaks is expected to lead to strong non-Markovian behavior.

In the biological situation, the FMO complex most likely obtains the excitation at sites
1, 6, or 8, since these BChls are close to the chlorosomal antennas, where photons
are absorbed. 
It is often assumed that this excitation is initially local to these sites.
%, although a delocalized excitation cannot be excluded. 
%
Low energy site 3 is the target site for the excitation and is close to the reaction center where further biochemical processes take place. 
In the ultrafast experiments, broad laser pulses excite a superposition of several delocalized exciton states. 
%

%%%%%%%%%%%%%%%%%%%%%%%%%%%%%%%%%%%%%%%%%%%%%%%%%%%%%%%%%%%%%%%%%%%%%%%%%%
%%                      The Analogue
%%%%%%%%%%%%%%%%%%%%%%%%%%%%%%%%%%%%%%%%%%%%%%%%%%%%%%%%%%%%%%%%%%%%%%%%%%
\section{The simulator}

It is challenging to simulate the open quantum system described in the previous section on 
conventional computers \cite{ishizaki2009,eisfeld2011,seibt2009}, 
even using modern parallel processing units \cite{strumpfer2009,kreisbeck2011,kreisbeck2012}.
Here we propose using flux qubits coupled with tunable flux-flux couplings for this task.
The environment is modeled by classical noise or quantum oscillators coupled to the flux qubits.
%

%%%%%%%%%%%%%%%%%%%%%%%%%%%%%%%%%%%%%%%%%%%%%%%%%%%%%%%%%%%%%%%%%%%%%%%%%%
%------------------------   F I G   2
\begin{figure}[t]
\resizebox{0.9\linewidth}{!}{\includegraphics{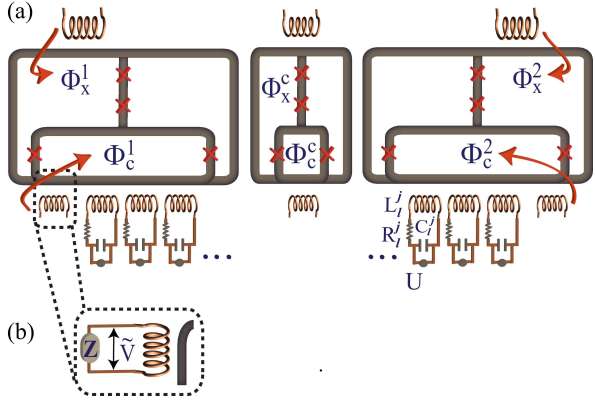}}
\caption{Circuit diagram of the proposed quantum simulator.
(a)~The qubit states are encoded in the quantized circulating current 
of the qubit loop.
The red crosses denote Josephson junctions.
Two flux qubits are coupled with a tunable $\sigma_z \sigma_z$-coupling.
Each of the qubits is independently coupled to a finite number of 
quantum LRC oscillators to simulate the non-Markovian vibrational environment.
(b)~Simulating the vibrational environment by adding a classical noise
to each qubit.
}\label{FMO-Simu}
\end{figure}
%%%%%%%%%%%%%%%%%%%%%%%%%%%%%%%%%%%%%%%%%%%%%%%%%%%%%%%%%%%%%%%%%%%%%%%%%%%

%%%%%%%%%%%%%%%%%%%%%%%%%%%%%%%%%%%%%%%%%%%%%%%%%%%%%%%%%%%%%%%%%%%%%%%%%%
\subsection{The system Hamiltonian}
Consider first a single flux qubit.
The relevant quantum states are the ones with magnetic flux pointing up  
$\ket{\uparrow}$ and down $\ket{\downarrow}$ or, equivalently, opposite directions of 
persistent current along the loop.
In this bare basis, the Hamiltonian of a flux qubit is given by 
\mbox{$H=  \, (\mathscr{E} \,\sigma_z + \Delta \,\sigma_x)/2$}, 
where $\mathscr{E}$ is the energy bias between $\ket{\uparrow}$ and 
$\ket{\downarrow}$ and $\Delta$ is the tunnel splitting 
between the two states.
Here $\mathscr{E} = 2 I_p(\phi_x - \phi_0/2) $  \cite{clarke} with $I_p$ being the persistent 
current of the qubit and $\phi_0=h/2e$  being the flux quantum.
$\mathscr{E}$ can be tuned to zero to neglect the \mbox{$\mathscr{E}\sigma _{z}$} term.

A tunable transverse interaction between flux qubits equivalent to that in 
$H_{\rm el}$ can be realized using additional  `coupler' qubits \cite{lloyd2007,ashhab2008}.
A schematic of such a simulator is given in Figure~\ref{FMO-Simu}~(a). 
The Hamiltonian of the coupled qubit system can be written as
\mbox{$
H_{\rm q}=  \,  \sum_{j=1}^N  \Delta_j \,\sigma_x^j /2
+  \sum_{i<j}^{N}\,  g_{ij}(\Delta^c_{ij}) \, \sigma_z^i\, \sigma_z^j 
$}
with \mbox{$g_{ij}(\Delta^c_{ij})$} being the coupling strength between flux qubits
$i$ and $j$,
which is given by 
\mbox{$g_{ij}(\Delta^c_{ij})\approx \mathscr{J}_{ij}-2\mathscr{J}_{ic}\mathscr{J}_{jc}/\delta_{ij}$},
where $\Delta^c_{ij}$ is the (tunable) tunnel splitting of the coupler qubit and 
we have defined \mbox{$\delta_{ij}\equiv\Delta_{ij}^c-(\Delta_i+\Delta_j)/2$} and
\mbox{$\mathscr{J}_{mn}\equiv \mathscr{M}_{mn}I_p^mI_p^n$} with ${m,n}\in{i,j,c}$ 
\cite{ashhab2008}.
Here, $\mathscr{M}_{mn}$ is the mutual inductance between qubits $m$ and $n$.
This expression is valid to leading order when 
$\delta_{ij}\gg|\Delta_i-\Delta_j|, \mathscr{J}_{ic}, \mathscr{J}_{jc}$.
Notice that by choosing the magnitude of $\Delta_{ij}^c$ to be smaller or larger than 
$(\Delta_i+\Delta_j)/2$ we can change the sign of the effective coupling.
Rewriting the above Hamiltonian in the energy eigenbasis of the qubit
\mbox{$\ket{\pm}=\left(\ket{\downarrow}+\ket{\uparrow}\right)/\sqrt{2}$}
converts \mbox{$\sigma_x^j\, \to \, \sigma_z^j$} and 
\mbox{$\sigma_z^i\, \sigma_z^j \, \to \, \sigma_x^i\, \sigma_x^j \approx
\left( \sigma_x^i \sigma_x^j + \sigma_y^i \sigma_y^j \right)/2$}
in the rotating wave approximation (neglecting strongly off-resonant couplings).
This results in 
\bea
\label{qubit_hamiltonian-02}
H_{\rm q} \approx \, \frac{1}{2}\,  \sum_{j=1}^N \Delta_j \,\sigma_z^j
+ \frac{1}{2} \sum_{i<j}^{N}\,  g_{ij}(\Delta^c_{ij}) \, 
\left( \sigma_x^i\, \sigma_x^j + \sigma_y^i\, \sigma_y^j \right)\, ,
\nn
\ea
which is of exactly the same form as the system part (first line) 
of Eq.~\ref{total_hamiltonian-02} with $\Delta_j$ and $ g_{ij}(\Delta^c_{ij})$ 
corresponding to $\tilde{\varepsilon }_{j}$ and  $V_{ij}$, respectively. 
It is advantageous for the experimental implementation to note that the dynamics of 
Eqs.~\ref{total_hamiltonian-02} and \ref{qubit_hamiltonian-02}
does not depend on absolute site energies $\tilde\varepsilon _{j}$ and $\Delta_j$ 
but only on energy differences $|\tilde\varepsilon _{i} - \tilde\varepsilon _{j}|$ and 
$|\Delta_i - \Delta_j|$, respectively.

\begin{figure}[t]
\centering
\resizebox{0.9\linewidth}{!}{\includegraphics{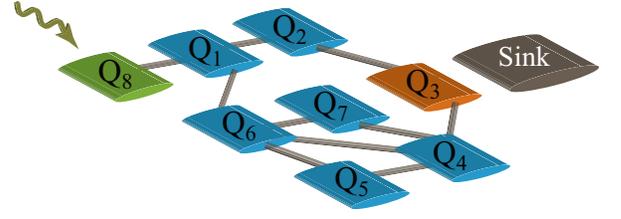}}
\caption{Experimental layout for simulating the exciton dynamics and  
environment assisted quantum transport (ENAQT) in the FMO complex 
(the architecture is based on the interactions given in Ref.~\cite{renger2011}, 
where for simplicity of the graphic the couplings below 15 cm$^{-1}$ are not shown).
$Q_i$ represent single flux qubits. 
To simulate a biologically relevant case, one of these qubits,  $Q_{8}$  
shown in green, is prepared initially in the excited state while the others are set to the ground state.
The measurement is performed on the target site, $Q_{3}$ shown in red.
Sinks can be used to trap the energy and quantify the transfer efficiency. }
\label{fig:ENAQT_Linear}
\end{figure}

The system of two coupled flux qubits  shown in Figure~\ref{FMO-Simu}~(a) can be 
extended to eight flux qubits with a special arrangement to simulate eight chlorophylls. 
An experimental layout simulating the electronic part of the FMO Hamiltonian
is given in Figure~\ref{fig:ENAQT_Linear}, where $Q_j$ represent flux qubits.
Static disorder can be simulated in our proposed scheme by varying the
tunnel splittings $\Delta_j$ in the flux qubits with each run of the experiment.
First all the qubits are in the ground state by simply allowing the system to relax.
Then they are initialized in a certain desired initial state to start the dynamics.
The excitation of a qubit is straightforward to achieve with the 
application of a resonant microwave excitation ($\pi$-pulse) carried by a microwave 
line which is connected to the respective qubit.
The technique has been used extensively, e.g., for the observation of 
Rabi oscillations in a flux qubit \cite{clarke,chiorescu}.
After some evolution time the populations  of the $\ket{\pm}$ states of the qubits are measured.
The measurement is initiated by applying a flux pulse to shift the qubit adiabatically away from 
$\mathscr{E}=0$ so that its eigenstates become largely $|\uparrow\rangle$ and $|\downarrow\rangle$ 
which can be distinguished via the flux induced in a nearby DC SQUID loop.
In addition, to capture the effect of reaction centers on the dynamics, we propose to add excitation sinks into
the superconducting circuit, see Figure~\ref{fig:ENAQT_Linear}.
This is done with additional terminated transmission lines or shunt resistors coupled to those sites
that are supposed to leak excitations.  

%%%%%%%%%%%%%%%%%%%%%%%%%%%%%%%%%%%%%%%%%%%%%%%%%%%%%%%%%%%%%%%%%%%%%%%%%%
%------------------------   F I G  3
%%%%%%%%%%%%%%%%%%%%%%%%%%%%%%%%%%%%%%%%%%%%%%%%%%%%%%%%%%%%%%%%%%%%%%%%%%

%%%%%%%%%%%%%%%%%%%%%%%%%%%%%%%%%%%%%%%%%%%%%%%%%%%%%%%%%%%%%%%%%%%%%%%%%%

\subsection{Engineering classical noise}

We will first discuss the more simple case of an environment
treated within the classical noise formalism of Eq.~\ref{classicHamil}.
We propose two schemes, an active approach and a passive one.
The basic idea of both approaches is to simulate the fluctuations of the 
transition energies by coupling classical noise to the qubits via the 
flux $\Phi_c^j$, see Figure~\ref{FMO-Simu}~(b). 
Such a noise affects the tunnel splitting $\Delta_{j}$ in the qubit Hamiltonian.
On one hand, the noise can be {\em actively} created and 
send to the qubit  by a time-dependent voltage $\tilde{V}$ applied to a control loop 
\cite{omelyanchouk2009}.
Regarding the {\em passive} approach, an environmental loop exhibits standard 
Johnson-Nyquist noise \cite{callen1951}.
The bath spectrum $J(\omega)$ of the noise seen by the qubit can be related to the
real part of the impedance $Z(\omega )$ \cite{wilhelm_nato2006}:
$ J(\omega )=K \,\omega \, \mathrm{Re}\left[Z(\omega)\right],$
with $K$ being a constant depending on the self-inductance of the
environment loop and its coupling strength to the flux qubit.
With classical circuit design $Z(\omega)$ can be tailored to produce the 
desired frequency dependence, for example, the ones in Refs. \cite{cho2005,adolphs2006,ishizaki2009} 

With simple classical noise a prototypical experiment of ENAQT 
can be performed \cite{patrick2009NJP}, see Figure~\ref{fig:ENAQT_Linear}.
In the FMO complex, the initial state of the simulation can be one of the sites $Q_1$, $Q_6$, or $Q_8$ which are
close to the antenna in the biological system.  
Measurement of success of the transport is performed at site $Q_3$ 
or by evaluating the population lost to the sink. 
Such an experiment can show that the environment is not always adversarial, but instead can
make certain processes, like quantum transport, more efficient 
(see the Supporting Information for more details).
This transport efficiency should exhibit a maximum at a dephasing rate that corresponds to 
room temperature in the biological system \cite{patrick2009NJP}.
Similar ideas of simulating ENAQT have been pursued in \cite{caruso2011,semiao2010}.
%

%%%%%%%%%%%%%%%%%%%%%%%%%%%%%%%%%%%%%%%%%%%%%%%%%%%%%%%%%%%%%%%%%%%%%%%%%%
%------------------------   F I G  4
\begin{figure}[t]
\centering
\resizebox{0.88\linewidth}{!}{\includegraphics{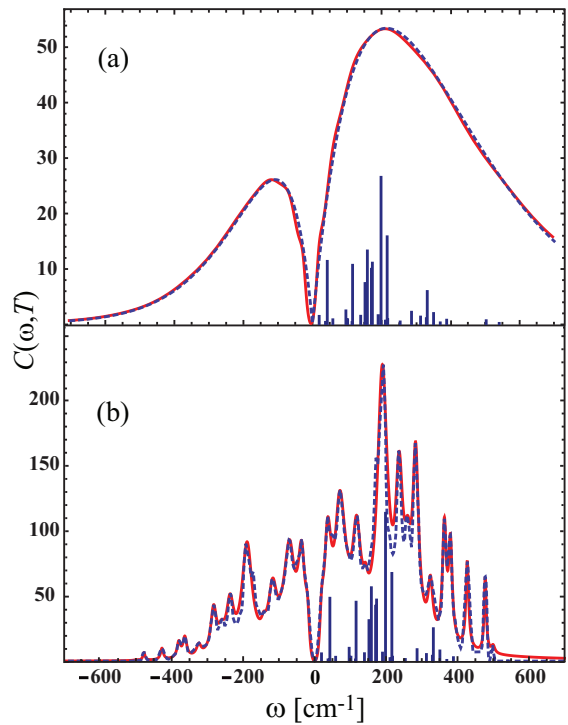}}
\caption{The parametrization of the temperature-dependent 
(a) super-Ohmic spectral density 
and 
(b) experimental \cite{wendling2000} spectral density of the FMO complex 
into distinct quantum oscillators.
The blue dashed lines are the spectral densities and the red solid lines are 
highly accurate simulation with LRC-oscillators coupled to the flux qubits, 
see Figure~\ref{FMO-Simu}~(a).
The temperature for both spectral densities is 300~K. 
The super-Ohmic spectral density is simulated with 6 damped LRC-Oscillators 
and the experimental one is simulated with 15 damped oscillators.
The blue bars show the transition energies \cite{adolphs2006} for the FMO complex.
The obtained parameters are given in the Supporting Information.
}
\label{fig:NMS300}
\end{figure}

%%%%%%%%%%%%%%%%%%%%%%%%%%%%%%%%%%%%%%%%%%%%%%%%%%%%%%%%%%%%%%%%%%%%%%%%%%

%%%%%%%%%%%%%%%%%%%%%%%%%%%%%%%%%%%%%%%%%%%%%%%%%%%%%%%%%%%%%%%%%%%%%%%%%%
%%         Simulation of FMO - Non-Markovian approach
%%%%%%%%%%%%%%%%%%%%%%%%%%%%%%%%%%%%%%%%%%%%%%%%%%%%%%%%%%%%%%%%%%%%%%%%%%
\subsection{Non-Markovian approach}
In order to simulate the complex environment described by Eq.~\ref{total_hamiltonian-02}
and capture the non-Markovian effects, we propose to couple each of the flux qubits 
inductively to an independent set of a few damped quantum LRC oscillators, 
see Figure~\ref{FMO-Simu}~(a).
The coupling Hamiltonian between the qubits and oscillators \cite{hauss2008}
in the energy eigenbasis is given by 
\mbox{$H_{\mathrm{q-osc}}=\sum_{j=1}^{N_{\rm LRC}}\sum_{k}\eta _{jk }\,\sigma
_{z}^{j}( {b_k^j}^{\dagger }+{b_k^j})$}, with ${b_k^j}^\dagger$ (${b_k^j}$) 
being the creation (annihilation) operator of the $k$-th oscillator in site $j$.
The coupling strength is given by \mbox{$\eta_{jk}\equiv \mathscr{M}_{jk} \,  I_{0}^{jk}\, (d\Delta_j/d\Phi_c^j)$},
where $I_0^k$ is the root mean square (RMS) amplitude of the current in the $k$-th oscillator ground state.

To simulate the original spectral density \ref{spectral_density}, 
we have to design the frequencies and couplings of the oscillators in such
a way that the spectral density is reproduced up to a global scaling factor.
From the implementation point of view, we are limited to a finite number of
oscillators.
Thus,  we decompose the spectral density of interest 
into a moderate number of spectral densities of damped oscillators.
The spectral density of a single oscillator coupled to a flux qubit can be derived  
by using a quantum Langevin equation approach \cite{bertet2005,bertet2005-2} and 
following the detailed balance condition \cite{egorov1999}
\bea
\label{LCSD}
C_{\rm osc}(\omega,T) = \mathfrak{D}
\left[\frac{e^{\hbar\omega/k_BT}}{\kappa^2+4(\omega-2\pi\omega_0)^2} 
+ \frac{1}{\kappa^2+4(\omega+2\pi\omega_0)^2}\right],
\nonumber
\ea
where
$\mathfrak{D} = (\sqrt{8/\pi}\,\kappa\, \eta^2)/(e^{\hbar\omega_0/k_BT}+1)$
with $\omega_0$ being the transition frequency of the oscillator and 
$\eta$ being the coupling strength of the oscillator to the flux qubit.
Here, \mbox{$\kappa=\kappa_0 \exp(-|\omega|/\alpha) \, \omega^2/\omega_0^2$} with 
$\kappa_0$ being the damping rate and $\alpha$ being a free parameter chosen 
reasonably to get the desired spectral density. 
By knowing the above spectral density for the damped quantum oscillators, 
we first simulate the temperature-dependent super-Ohmic spectral density.
At 300~K this spectral density can be simulated with a set of 6 
LRC oscillators coupled to each of the flux qubits, see Figure~\ref{fig:NMS300}~(a), 
and at 77~K it can be simulated with a set of 7 oscillators (see Figure~S1~(a) in the Supporting Information).
For the experimental spectral density, we need to couple, for example, 15 oscillators to each qubit, 
see Figure~\ref{fig:NMS300}~(b). 
Notice that the  so-obtained SDs  in Figure~\ref{fig:NMS300} are highly accurate and
we can use fewer coupled oscillators if we are interested in less details of the spectral densities. 
The coupling of the oscillators to the flux qubit results in an additional shift
of the qubit tunnel splitting $\Delta_j$ due to the reorganization energy of the oscillators. 
This has to be taken into account  in the design of the circuit energy landscape.

%%%%%%%%%%%%%%%%%%%%%%%%%%%%%%%%%%%%%%%%%%%%%%%%%%%%%%%%%%%%%%%%%%%%%%%%%%
%%                    Experimental Parameters
%%%%%%%%%%%%%%%%%%%%%%%%%%%%%%%%%%%%%%%%%%%%%%%%%%%%%%%%%%%%%%%%%%%%%%%%%%
%
\section{Experimental feasibility}
The simulation of the time evolution of the FMO complex requires a moderately coherent eight-qubit system,
which would be realizable using the flux qubits demonstrated in Ref.~\cite{paauw2009} and the coupling 
geometries and methods of Refs.~\cite{lloyd2007,ashhab2008}.
In the FMO complex the site energies/chlorophyll excitation energies
are around \mbox{$12500\,\,\mathrm{cm}^{-1}$}, 
with the average site-dependent static shifts of the order of \mbox{$250\,\,\mathrm{cm}^{-1}$}. 
We emphasize again that only the {\em site-energy differences},  not the site energies themselves,
play  a  role in the single-exciton dynamics. 
The magnitudes of the coupling strengths between the chlorophyll molecules are smaller
than  \mbox{$120\,\, \mathrm{cm}^{-1}$} \cite{adolphs2006}. 
For superconducting flux qubits, implementable range of the tunnel bias $\Delta_j$ is in 
the range of approximately zero to 13~GHz \cite{paauw2009}, while the coupling strengths $g_{ij}$ 
were measured in the range of  one GHz to approximately zero \cite{lloyd2007}.
%

%%%%%%%%%%%%%%%%%%%%%%%%%%%%%%%%%%%%%%%%%%%%%%%%%%%%%%%%%%%%%%%%%%%%%%%%%%
\begin{table*}
\begin{tabular}{lll}
 {\bf Parameter} & {\bf FMO model} & {\bf Quantum simulator}\\
\hline 
decay time ($\mathbb{T}_1$) & $\approx$ ns & $\approx$ 10 $\mu$s\\
average exciton transfer time & $\approx$ 5~ps & $\approx$ 25 ns \\
decay time between the exciton states & $\approx$ ps & $\approx$ 5 ns \\
dephasing in exciton manifold & $\approx$ 100 fs & $\approx$ 500 ps \\
time scale of quantum beatings ($\tau_{\rm osc}$)  & $\approx$ 200 fs & $\approx$ 1 ns \\
coupling between sites & $\approx$ 10 cm$^{-1}$ - 122 cm$^{-1}$  & $\approx$ 60 MHz - 730 MHz\\
relative static site energy shifts & $\approx$ 10 cm$^{-1}$ - 500 cm$^{-1}$ &  $\approx$ 60 MHz - 3 GHz\\
temperature &  300 K& 60 mK \\
\hline 
\end{tabular}
\caption{\label{Exp-Para}
Comparison of parameters for the FMO complex and the quantum simulator.
The timescales shown below are for the dressed states of flux qubits coupled to the 
quantum harmonic oscillators. 
For more details see Figure~S3 and Tabels~S1, S2, and S3 in the Supporting Information.
Notice that the decay 
time in a single qubit ($\mathbb{T}_1$) does not need to be mapped directly from the FMO dynamics.
With nowadays achievable decay times in superconducting qubits, which are 3 orders of 
magnitude larger than the excitation transfer time between the qubits,  the dynamics of the FMO complex can be simulated.}
\end{table*}

%%%%%%%%%%%%%%%%%%%%%%%%%%%%%%%%%%%%%%%%%%%%%%%%%%%%%%%%%%%%%%%%%%%%%%%%%%

The parameters of the proposed quantum simulator are scaled through 
the time scale of quantum beatings ($\tau_{\rm osc}$) to be 
consistent internally as well as  with the implementation restrictions,  
see Table~\ref{Exp-Para}.
Photosynthesis occurs at ambient temperatures, e.g., 300~K,
which then maps to 60~mK in superconducting-circuit experiments.
The FMO dynamics is usually considered for up to 5 ps, which translates 
to the timescale of 25 ns in the flux qubits.
The energy relaxation time ($\mathbb{T}_1$) of a single qubit has been found 
\cite{chiorescu,coherence-time} to be on the order of a few $\mu$s, being a few orders of
magnitude larger than the required exciton transfer time. 
Several coherent beatings between two coupled qubits have been observed  \cite{lloyd2007}.
Table~\ref{Exp-Para} represents the summary of the  required range of parameters for the
superconducting simulator to imitate the dynamics of the FMO complex.

In simulating the quantum environment of the FMO complex, 
the required transition frequency of the LRC oscillators 
are in the range of 120~MHz to 3~GHz and the coupling to the qubits are 
around 8~MHz to 100~MHz, see Tables S2 and S3 in the Supporting Information.
These parameters are experimentally reasonable.
The geometry of Figure~\ref{fig:ENAQT_Linear} may cause
space problems in implementing this system. 
To avoid this, the parallel combination of resonators used to implement the desired spectral density 
can be mapped, for example,  to a linear chain of oscillators  \cite{mori1965,hughes2009,prior2010}, 
such that only a single resonator would need to be coupled directly to each qubit, 
see Figure~S2 in the Supporting Information. 
The quality factor (transition frequency/bandwidth) of the quantum oscillators in our proposed 
simulator are up to 50 or less (i.e., each of the oscillators is strongly overcoupled to an output 50~Ohm line).
The coupling to the output line can be tunable to adjust the quality factor
of each oscillator in situ.
Finally, low-frequency flux noise (1/$f$ noise) is one
of the decoherence sources in flux qubits.
In our proposal, this noise is suppressed at the optimal working point \cite{yoshihara2006}, where  $\mathscr{E} =0$.
All of the above numbers and observations suggest that the site energy differences to coupling ratios of
the FMO complex as well as corresponding temperature and environmental couplings are achievable 
with superconducting circuits.
%
%%%%%%%%%%%%%%%%%%%%%%%%%%%%%%%%%%%%%%%%%%%%%%%%%%%%%%%%%%%%%%%%%%%%%%%%%%
%%                               Conclusion
%%%%%%%%%%%%%%%%%%%%%%%%%%%%%%%%%%%%%%%%%%%%%%%%%%%%%%%%%%%%%%%%%%%%%%%%%%
%
\section{Conclusion}

We have demonstrated that an appropriately designed network of superconducting qubit-resonator design can simulate 
not only the coherent exciton transport in photosynthetic complexes, but also the 
effect of a complicated quantum environment.
We have highlighted its experimental feasibility with present-day technology. 
In particular, we have shown that a straightforward combination of superconducting 
qubits (representing the chlorophyll molecules) and 
resonators (simulating the phonon environment) can be used to obtain a reasonable 
approximation to the exciton and phonon degrees of freedom in the FMO complex.
For example, we show ways to engineer a spectral density that
reproduces the one of the biological system.
One of the advantages of our proposed quantum simulator, compared to the computational 
methods, is simulating both diagonal and off-diagonal noise. 
Because of the additional complexity of considering the off-diagonal noise 
most of the non-Markovian computational methods only take the diagonal noise into the account. 
Another advantage is that, by design, we have a single molecule setup while all the ultrafast 
experiments use an ensemble of light harvesting complexes. 
This allows for more detailed studies of non-Markovian energy transfer pathways.

An important feature of our proposal is the potential to achieve a high level of 
environment engineering, in such a way that external noise is used to benefit 
the quantum coherent energy transfer process inside the molecule.  
However, the broader scope of our work is along the lines of \emph{biomimesis}: 
the artificial recreation of biological processes, which are already 
highly optimized through evolution.
%
%%%%%%%%%%%%%%%%%%%%%%%%%%%%%%%%%%%%%%%%%%%%%%%%%%%%%%%%%%%%%%%%%%%%%%%%%%
%%%                       Acknowledgments
%%%%%%%%%%%%%%%%%%%%%%%%%%%%%%%%%%%%%%%%%%%%%%%%%%%%%%%%%%%%%%%%%%%%%%%%%%
\begin{acknowledgments}
The authors are indebted to S.~Ashhab, M.~Geller, F.~Nori, S.~Valleau,  S.~Huelga, 
and M.~H.~S.~Amin for valuable conversations.  
We acknowledge DARPA Grant No.~N66001-10-1-4063, NIST Award No.~60NANB10D267 and
DTRA Grant No.~HDTRA1-10-1-0046.
This material is based upon work supported as a part of the Center for 
Excitonics, as an Energy Frontier Research Center funded by the U.S. Department of 
Energy, Office of Science, Office of Basic Energy Sciences under Award 
number DESC0001088.
AE acknowledges financial support from the DFG under Contract No.~Ei~872/1-1.
The work at MIT Lincoln Laboratory is sponsored by the United States Air Force 
under Air Force Contract No.~FA8721-05-C-0002. 
Opinions, interpretations, recommendations and conclusions are those of the 
authors and are not necessarily endorsed by the United States Government.
\end{acknowledgments}

$^*$\,{\footnotesize\sf aspuru@chemistry.harvard.edu}\,\,  
%%%%%%%%%%%%%%%%%%%%%%%%%%%%%%%%%%%%%%%%%%%%%%%%%%%%%%%%%%%%%%%%%%%%%%%%%%
%%%%%%%%%%%%%%%%%%%%%%%%%%%%%%%%%%%%%%%%%%%%%%%%%%%%%%%%%%%%%%%%%%%%%%%%%%
%%%%%%%%%%%%%%%%%%%%%%%%%%%%%%%%%%%%%%%%%%%%%%%%%%%%%%%%%%%%%%%%%%%%%%%%%%

%%%%%%%%%%%%%%%%%%%%%%%%%%%%%%%%%%%%%%%%%%%%%%%%%%%%%%%%%%%%%%%%%%%%%%%%%%
%%                    Bibliography
%%%%%%%%%%%%%%%%%%%%%%%%%%%%%%%%%%%%%%%%%%%%%%%%%%%%%%%%%%%%%%%%%%%%%%%%%%
%----------------------------------------------------------

%%%%%%%%%%%%%%%%%%%%%%%%%%%%%%%%%%%%%%%%%%%%%%%%%%%%%%%%%%%%%%%%%%%%%%%%%%\

\end{document}

\section{Supporting Information}

%Mostame, Rebentrost,  Eisfeld, Kerman, Tsomokos and Aspuru-Guzik

%%%%%%%%%%%%%%%%%%%%%%%%%%%%%%%%%%%%%%%%%%%%%%%%%%%%%%%%%%%%%%%%%% 

\section{Coordinate representation of the model Hamiltonian}
In this section we derive the model Hamiltonian of a single molecule (Eq.~[{\bf 1}] in the main article for $N=1$)
 from the Born-Oppenheimer approximation.
Restricting to two electronic states, the ground state $|g\rangle$ and the excited state $|e\rangle$, the
Hamiltonian in the Born-Oppenheimer approximation is \cite{may-kuehn}:
\bea
H = H_{\rm nuc,g}(R) |g\rangle \langle g | + H_{\rm nuc,e}(R) |e\rangle \langle e | \, ,
\ea
where $R$ describes the collection of $3N_{\rm nuc}$ modes relevant to the molecule
(both local and protein modes), \mbox{$R=\{R_1,\cdots,R_{3N_{\rm nuc}}\}$}, 
with $N_{\rm nuc}$ being the number of nuclei.
The Hamiltonians $H_{\rm nuc,g/e}(R)$ describe the kinetic and potential
energy of the nuclei, $T_{\rm nuc}$ and $V_{\rm nuc}$, respectively:
\mbox{$H_{\rm nuc,g/e}(R) = T_{\rm nuc}+V_{\rm nuc,g/e}(R)$}.
The potential energy is given by 
$V_{\rm nuc,g/e}(R)=V_{\rm nuc-nuc}(R)+E_{\rm g/e}(R)$ 
with the inter-nuclear potential energy 
$V_{\rm nuc-nuc}$ and the potential energy due to the electrons
$E_a(R)$. 
We assume a displaced harmonic oscillator model for
the potential of ground and excited state:
\begin{eqnarray}
V_{\rm nuc,g}(q) &=&U_g + \sum_{i=1}^{3N_{\rm nuc}} \frac{\hbar \omega_i}{2} q_i^2, \\
V_{\rm nuc,e}(q) &=&U_e + \sum_{i=1}^{3N_{\rm nuc}} \frac{\hbar \omega_i}{2} (q_i-d_i)^2.
\end{eqnarray}
Here, we introduced the renormalized coordinates $q=R-R_0$, where
$R_0$ are the equilibrium positions in the electronic ground state
(minimum of the ground state potential energy surface). 
The respective energies of the electronic states at the minimum of the respective potentials are $U_g$ and
$U_e$. 
We have assumed that the frequency $\omega_i$ of mode $i$ 
remains unchanged in the exited state. 
The displacement of the $i$th mode in the excited state is given by $d_i$.  
The $q$-dependent energy gap is given by the difference of the two potentials:
\begin{equation}
V_{\rm nuc,e}(q)-V_{\rm nuc,g}(q)= \Delta U + \sum_{i=1}^{3N_{\rm nuc}} 
\frac{\hbar \omega_i}{2} d_i^2 - \sum_{i=1}^{3N_{\rm nuc}} \hbar\omega_i d_i q_i\, ,
\nonumber
\end{equation}
where the first term $\Delta U=U_e - U_g$ is the energy
difference between the potential minima of ground and excited state. 
The second term gives the reorganization energy:
\begin{equation} 
\lambda = \sum_i \lambda_i =\sum_i \frac{\hbar \omega_i}{2} d_i^2 \, .
\end{equation} 
The third term gives the linear dependence of the gap on the coordinates of the harmonic oscillator,
and is the exciton-vibrational coupling term. 
The total Hamiltonian in the harmonic approximation is thus:
\begin{eqnarray}
H_{\rm tot} &= & \underbrace{ \left( \Delta U + \sum_i \lambda_i \right ) |e\rangle \langle e |}_{H_{\rm el}}  
\nn
&& + \underbrace{ \left( T_{\rm nuc} + \sum_i \frac{\hbar \omega_i}{2} q_i^2 \right)\mathbf{1} }_{H_{\rm ph}} 
\nn
&& + \underbrace{\sum_i \hbar\omega_i d_i q_i |e\rangle \langle e |}_{H_{\rm el-ph}}.
\end{eqnarray}
Here we defined the respective Hamiltonians for the electronic system, phonon bath and
electron-phonon coupling, $H_{\rm el}$, $H_{\rm ph}$, and $H_{\rm el-ph}$. 
The system identity operator is given by $\mathbf{1}=|g\rangle \langle g|+ |e\rangle \langle e|$.
%%%%%%%%%%%%%%%%%%%%%%%%%%%%%%%%%%%%%%%%%%%%%%%%%%%%%%%%%%%%%%%%%%%%%%%%%%
%------------------------   F I G  5
\begin{figure}[t]
\centering
\resizebox{0.8\linewidth}{!}{\includegraphics{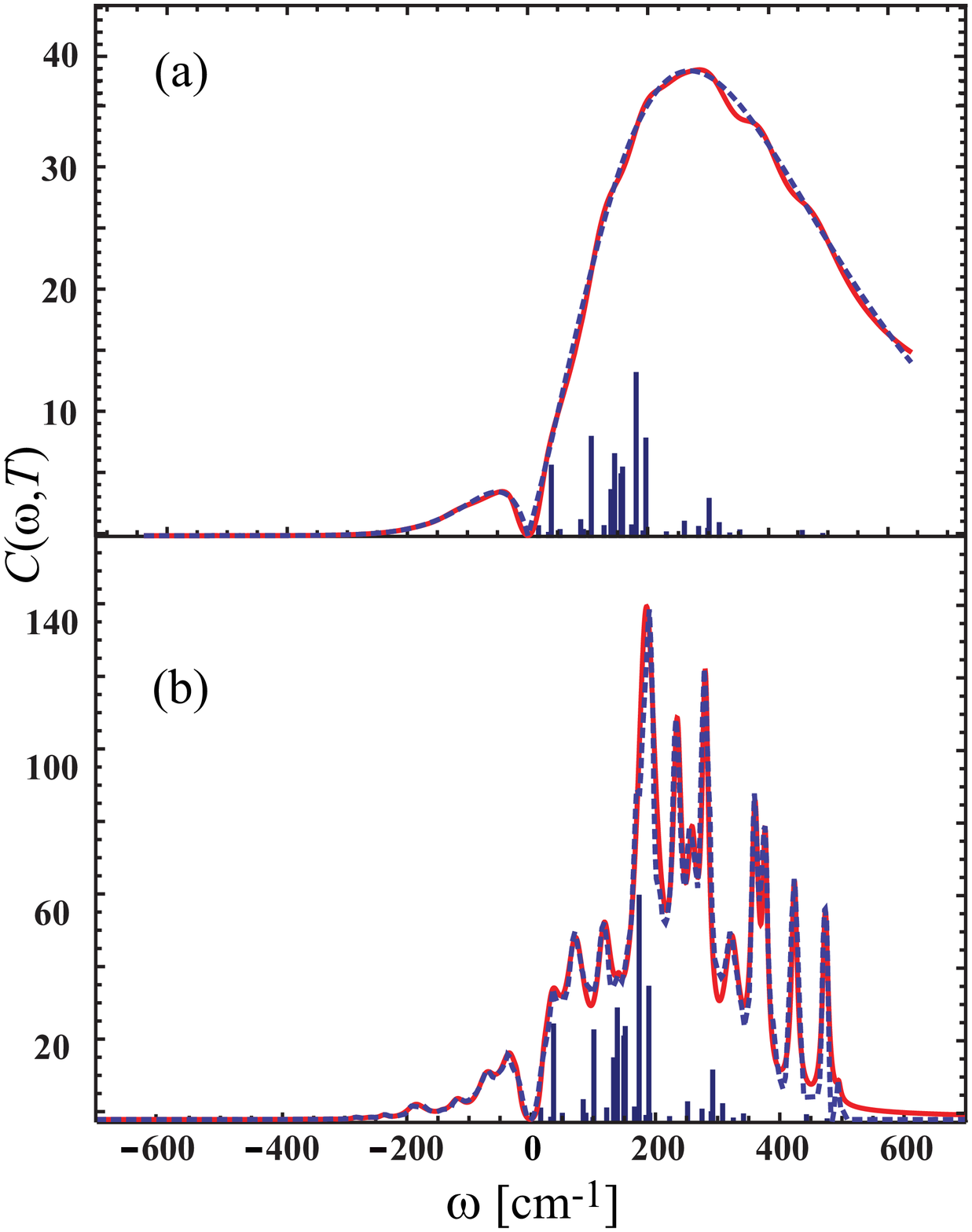}}
\caption{The parametrization of the temperature-dependent 
(a) super-Ohmic mode density 
and 
(b) experimental \cite{wendling2000} mode density of the FMO complex 
into distinct quantum oscillators.
The blue dashed lines are the mode densities and the red solid lines are accurate  
simulation with LRC-oscillators coupled to the flux qubits, see Figure~1~(a) 
of the main article.
The temperature for both mode densities is 77~[K]. 
The super-Ohmic mode density is simulated 
with 7 damped LRC-oscillators and the experimental one is simulated with 15 
damped oscillators.
The blue bars show the transition energies \cite{adolphs2006} for the FMO complex. 
}
\label{fig:NMS77}
\end{figure}
%%%%%%%%%%%%%%%%%%%%%%%%%%%%%%%%%%%%%%%%%%%%%%%%%%%%%%%%%%%%%%%%%%%%%%%%%%

%%%%%%%%%%%%%%%%%%%%%%%%%%%%%%%%%%%%%%%%%%%%%%%%%%%%%%%%%%%%%%%%%%%%%%%%%%

\section{Energy transfer pathways for the FMO complex }
We briefly explain the red arrows in Figure~1 (b), which show schematically the 
downwards energy transfer pathways for the FMO complex, similar to Ref.~\cite{brixner2005} 
which considered seven-site model for the FMO complex. 

From a system-bath model like Eq.~(1) (main article), one can derive a master equation 
for the density matrix by using, for example, Redfield theory with the secular approximation \cite{breuer}.
Redfield theory assumes weak coupling and a Markovian bath. 
This leads to decoherence rates in the energy basis between energy states 
$M$ and $N$ given by \cite{adolphs2006} (without loss of generality $\hbar \omega_{MN} = E_M - E_N >0$):

\begin{eqnarray}
\Gamma_{MN}^{\uparrow} &=& 2 \pi \, \gamma_{MN} J(\omega_{MN})\,  n(\omega_{MN} )\, ,
\\
\Gamma_{MN}^{\downarrow} &=& 2 \pi \,  \gamma_{MN} J(\omega_{MN}) 
\left[n(\omega_{MN} ) + 1 \right]\, .
\end{eqnarray}
Here, $\Gamma_{MN}^{\uparrow}$ ($\Gamma_{MN}^{\downarrow}$) is the rate up (down) in energy
and $n(\omega_{MN} )$ is the mean number of vibrational quanta with energy 
$\hbar \omega_{MN} $ that are excited at a given temperature $T$:
\bea
n(\omega_{MN} )=1/[\exp(\hbar \omega_{MN}/k_BT)-1]\, .
\ea
The factor \mbox{$\gamma_{MN} = \sum_j |\bracket{M}{j}|^2|\bracket{j}{N}|^2 $}
arises from the basis transformation between site and energy basis.
In Figure~1~(a), the red arrows show a selected number of downward transitions with 
\mbox{$\gamma_{MN} J(\omega_{MN}) \ge 0.3$~cm$^{-1}$}.

%%%%%%%%%%%%%%%%%%%%%%%%%%%%%%%%%%%%%%%%%%%%%%%%%%%%%%%%%%%%%%%%%%%%%%%%%%
%------------------------   F I G  6
\begin{figure}[t]
\begin{center}
\resizebox{0.98\linewidth}{!}{\includegraphics{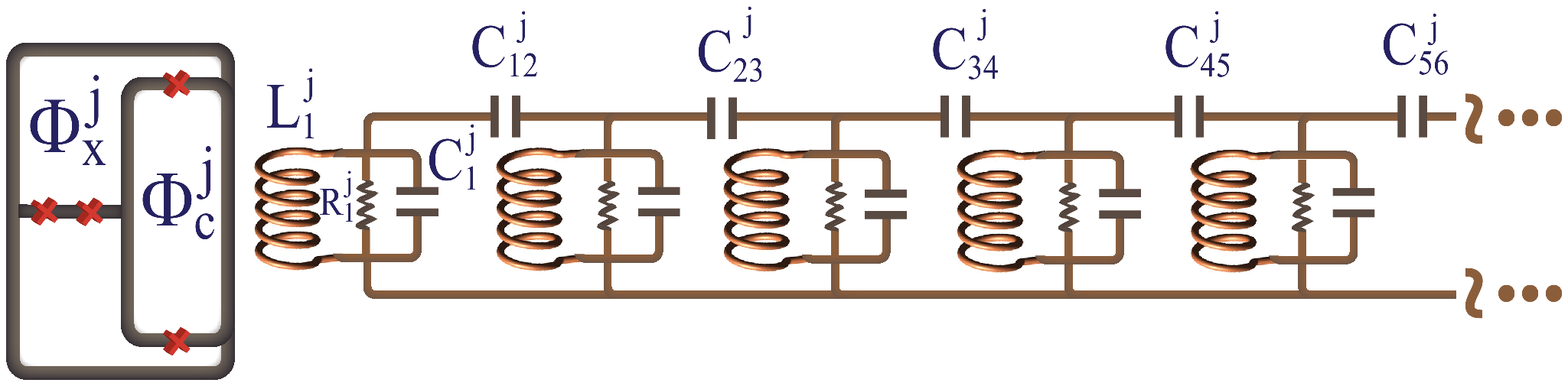}}
\caption{The parallel combination of LRC oscillators shown in Figure.~2~(a) of the main article 
can be mapped to a linear chain  of oscillators, such that only a single resonator 
would need to be coupled directly to each qubit.
The oscillators can be coupled, for example, capacitively via capacitors $C_{mn}^j$.
}
\end{center}
\label{fig:NMS77}
\end{figure}
%%%%%%%%%%%%%%%%%%%%%%%%%%%%%%%%%%%%%%%%%%%%%%%%%%%%%%%%%%%%%%%%%%%%%%%%%%
%%%%%%%%%%%%%%%%%%%%%%%%%%%%%%%%%%%%%%%%%%%%%%%%%%%%%%%%%%%%%%%%%% 
\section{Environment-assisted quantum transport}

Here we briefly discuss the main features of environment-assisted quantum transport and 
what is to be expected from an experiment scanning the ratio of dephasing rate over system energy scale. 
For the qubit system in our proposed quantum simulator, the couplings and the differences 
in the qubit splittings give a general energy scale $\Lambda $. 
The site energy level fluctuations lead to pure dephasing as the dominant decoherence mechanism,
which is phenomenologically characterized by a pure dephasing rate $\gamma$. 
In the active noise engineering case,  each site is driven, for example, 
by white noise with an amplitude $\sqrt{\gamma}$. 
The amplitude can be easily tuned in the external noise generator.
In the passive case, the noise level can be regulated by the temperature of the sample.
For both cases, if the dephasing rate $\gamma $ is much smaller than the energy scale $\Lambda $, 
quantum localization is predicted to arise from the disorder in the energy levels.
This leads to a small population at the target site. 
Increasing the dephasing rate such that $\gamma \approx \Lambda $
is expected \cite{patrick2009NJP} to lead to an increased population at the target site.
Finally, it is expected that for the dephasing rate $\gamma \gg \Lambda $ 
diminished population arrives at the target site, since quantum transport is suppressed by 
the Zeno effect.
%
%%%%%%%%%%%%%%%%%%%%%%%%%%%%%%%%%%%%%%%%%%%%%%%%%%%%%%%%%%%%%%%%%%%%%%%%%%
%------------------------   F I G  7
\begin{figure}[h]
\centering
\resizebox{0.95\linewidth}{!}{\includegraphics{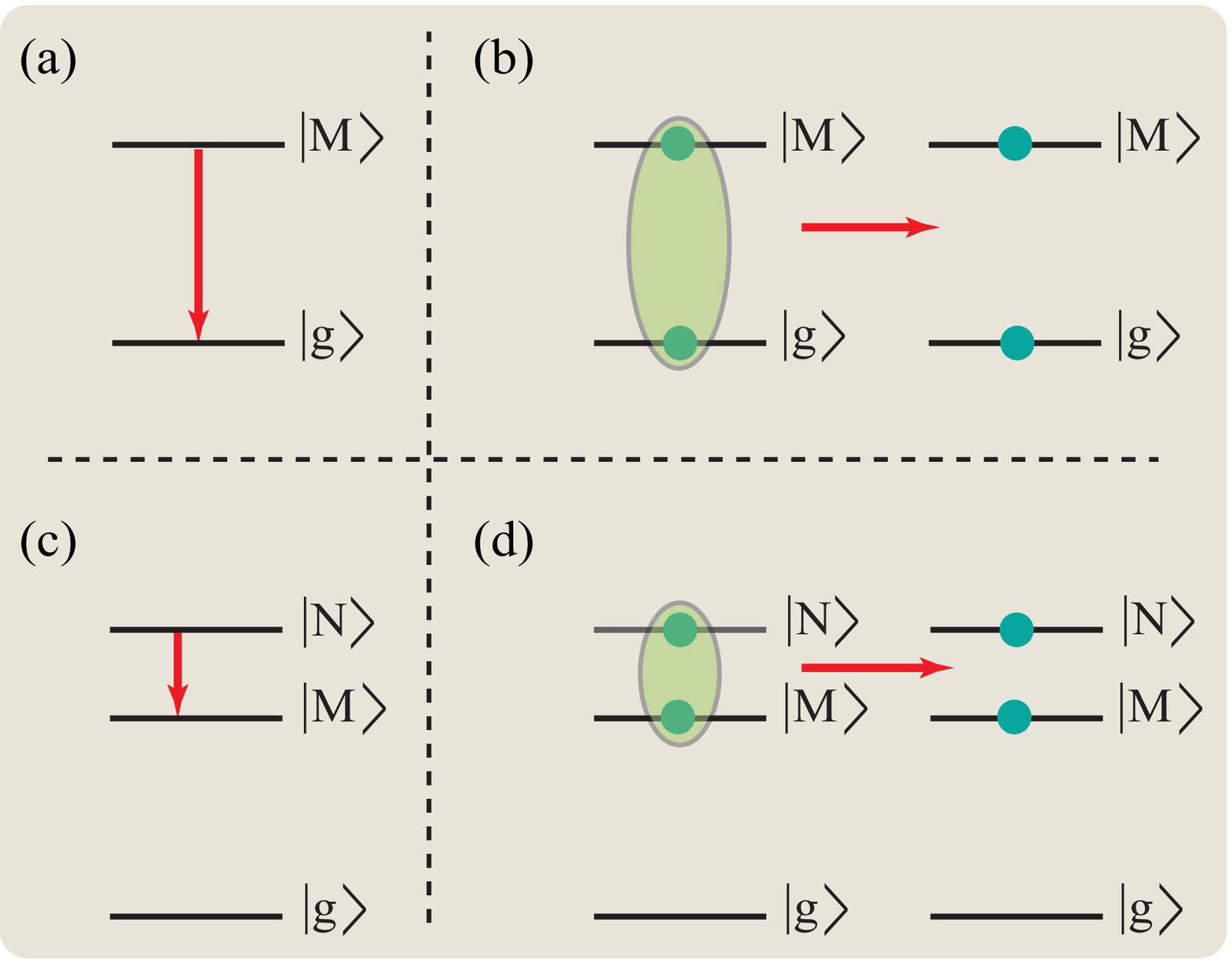}}
\caption{Sketch of the basic processes in excitonic energy transfer. 
Confer to Table~1 in the main text (or Table~S1) for numerical values of 
the time scales of the respective processes. 
Let $|g\rangle$ be the electronic ground state and $|M\rangle$ and $|N\rangle$ 
be two delocalized electronic excited states. 
(a) Decay between an excited state and the ground state,
characterized with the \textit{decay time} $\mathbb{T}_1$ (Another name for
this process is exciton recombination). 
(b) Dephasing of a superposition between ground state and an excited state. 
This process usually happens on a very fast time scale and is not relevant for the
present discussion. 
(c) Decay in the single exciton manifold without the loss of the excitation 
to the ground state, characterized by the \textit{decay time between exciton states}. 
(d) Dephasing in the single exciton manifold. 
Consider a superposition of exciton states $|\psi\rangle=\frac{1}{\sqrt{2}}(|M\rangle + |N\rangle)$. 
Then this dephasing process causes the initial density matrix $|\psi\rangle \langle \psi|$ 
to decay to an equal mixture $1/2(|M\rangle \langle M| + |N\rangle \langle N|)$ at long times.
The time scale of this process is characterized by the 
\textit{dephasing time in the single exciton manifold}. 
}
\label{fig:NMS77}
\end{figure}
%%%%%%%%%%%%%%%%%%%%%%%%%%%%%%%%%%%%%%%%%%%%%%%%%%%%%%%%%%%%%%%%%%%%%%%%%%

%%%%%%%%%%%%%%%%%%%%%%%%%%%%%%%%%%%%%%%%%%%%%%%%%%%%%%%%%%%%%%%%%%%%%%%%%%
%%                    Bibliography
%%%%%%%%%%%%%%%%%%%%%%%%%%%%%%%%%%%%%%%%%%%%%%%%%%%%%%%%%%%%%%%%%%%%%%%%%%
%----------------------------------------------------------

%----------------------------------------------------------

%%%%%%%%%%%%%%%%%%%%%%%%%%%%%%%%%%%%%%%%%%%%%%%%%%%%%%%%%%%%%%%%%%%%%%%%%%
%                         Table 1
%%%%%%%%%%%%%%%%%%%%%%%%%%%%%%%%%%%%%%%%%%%%%%%%%%%%%%%%%%%%%%%%%%%%%%%%%%
\begin{table*}[h]
\begin{tabular}{lll}
 \textbf{Parameter} & \textbf{FMO model} & \textbf{Quantum simulator}\\
\hline \\
decay time ($\mathbb{T}_1$) & $\approx$ ns & $\approx$ 10 $\mu$s\\
(single site electron-hole recombination) & & \\ \hline \\
average exciton transfer time & $\approx$ 5 ps & $\approx$ 25 ns \\
(from site 8 to site 3) & & \\ \hline \\
decay time between the exciton states & $\approx$ ps & $\approx$ 5 ns \\
(jump between exciton states) & & \\ \hline \\
dephasing in exciton manifold & $\approx$ 100 fs & $\approx$ 500 ps \\
(pure dephasing) & & \\ \hline \\
time scale of quantum beatings ($\tau_{\rm osc}$)  & $\approx$ 200 fs & $\approx$ 1 ns \\
 \hline \\
coupling between sites & $\approx$ 10 cm$^{-1}$ - 122 cm$^{-1}$   
& $\approx$ 60 MHz - 730 MHz \\
\hline \\
relative static site energy shifts & $\approx$ 10 cm$^{-1}$ - 500 cm$^{-1}$
 &  $\approx$ 60 MHz - 3 GHz\\
$\left| \tilde{\varepsilon_i} - \tilde{\varepsilon_j }\right| 
\equiv \left| \Delta_i - \Delta_j \right|$ & &  \\ \hline \\
dynamic fast fluctuations \cite{engel2010} & $\approx$ 250 $\pm$ 100 cm$^{-1}$ at 300 K
&  $\approx$ 1.5 GHz $\pm$ 600 MHz\\
(dephasing rate) & $\approx$ 40 $\pm$ 10 cm$^{-1}$ at 77 K
&  $\approx$ 240 MHz $\pm$ 60 MHz  \\ \hline \\
&  300 K $\approx$ 208 cm$^{-1}$ &  60 mk $\approx$ 1.2 GHz \\
temperature &  100 K $\approx$ 69.5 cm$^{-1}$ &  20 mk $\approx$ 417 MHz\\
 &  77 K $\approx$ 53 cm$^{-1}$ & 15 mk $\approx$ 317 MHz  \\ \hline
\end{tabular}
\caption{\label{Exp-Para21}
Comparison of parameters for the FMO complex and the quantum simulator.
The timescales shown below are for the dressed states of flux qubits coupled to the 
quantum harmonic oscillators.
Notice that the decay 
time in a single qubit ($\mathbb{T}_1$) does not need to be mapped directly from the FMO dynamics.
With nowadays achievable decay times in superconducting qubits, which are 3 orders of 
magnitude larger than the excitation transfer time between the qubits, the dynamics of the FMO complex can be simulated.
}
\end{table*}

%%%%%%%%%%%%%%%%%%%%%%%%%%%%%%%%%%%%%%%%%%%%%%%%%%%%%%%%%%%%%%%%%%%%%%%%%%
%                         Table 2
%%%%%%%%%%%%%%%%%%%%%%%%%%%%%%%%%%%%%%%%%%%%%%%%%%%%%%%%%%%%%%%%%%%%%%%%%%
\begin{table*}[h]
\begin{tabular}{@{\vrule height 10.5pt depth4pt  width0pt}lccc}
&\multicolumn1c{\hspace{3cm} \bf{FMO complex}}\\
\noalign{\vskip-11pt}
\\
\hline 
&\textbf{transition frequency} &\textbf{coupling strength}&\textbf{quality factor} \\
\textbf{oscillator~No.~1}& $\approx$ 27 cm$^{-1}$ & $\approx$ 2.42 cm$^{-1}$ & $\approx$ 0.67 \\
\textbf{oscillator~No.~2}& $\approx$ 74 cm$^{-1}$ & $\approx$ 8.60 cm$^{-1}$ & $\approx$ 0.49 \\
\textbf{oscillator~No.~3}& $\approx$ 140 cm$^{-1}$ & $\approx$ 11.98 cm$^{-1}$ & $\approx$ 0.47 \\
\textbf{oscillator~No.~4}& $\approx$ 246 cm$^{-1}$ & $\approx$ 14.10 cm$^{-1}$ & $\approx$ 0.80 \\
\textbf{oscillator~No.~5}& $\approx$ 380 cm$^{-1}$ & $\approx$ 10.00 cm$^{-1}$ & $\approx$ 1.27 \\
\textbf{oscillator~No.~6}& $\approx$ 560 cm$^{-1}$ & $\approx$ 5.40 cm$^{-1}$ & $\approx$ 1.84 \\
\hline
 \\
&\multicolumn1c{\hspace{3cm} \bf{Quantum simulator}}\\
\noalign{\vskip-11pt}
\\
\hline 
&\textbf{transition frequency} &\textbf{coupling strength}&\textbf{quality factor} \\
\textbf{oscillator~No.~1}& $\approx$ 162 MHz & $\approx$ 14.50 MHz & $\approx$ 0.67 \\
\textbf{oscillator~No.~2}& $\approx$ 444 MHz & $\approx$ 51.56 MHz & $\approx$ 0.49 \\
\textbf{oscillator~No.~3}& $\approx$ 839 MHz & $\approx$ 71.83 MHz & $\approx$ 0.47 \\
\textbf{oscillator~No.~4}& $\approx$ 1.5 GHz & $\approx$ 84.54 MHz & $\approx$ 0.80 \\
\textbf{oscillator~No.~5}& $\approx$ 2 GHz & $\approx$  59.95 MHz & $\approx$ 1.27 \\
\textbf{oscillator~No.~6}& $\approx$ 3 GHz & $\approx$ 32.38 MHz & $\approx$ 1.84 \\
\hline
\end{tabular}
\caption{\label{Exp-Para22}Decomposition of the temperature-dependent super-Ohmic mode density
at 300~K shown in Figure~4~(a) of the main article and simulation with 6 LRC-oscillators coupled to 
each flux qubit, see Figure~2~(a) of the main article.}

\end{table*}

%%%%%%%%%%%%%%%%%%%%%%%%%%%%%%%%%%%%%%%%%%%%%%%%%%%%%%%%%%%%%%%%%%%%%%%%%%
%                         Table 3
%%%%%%%%%%%%%%%%%%%%%%%%%%%%%%%%%%%%%%%%%%%%%%%%%%%%%%%%%%%%%%%%%%%%%%%%%%
\begin{table*}[h]
\begin{tabular}{@{\vrule height 10.5pt depth4pt  width0pt}lccc}
&\multicolumn1c{\hspace{3cm} \bf{FMO complex}}\\
\noalign{\vskip-11pt}
\\
\hline 
&\textbf{transition frequency} &\textbf{coupling strength}&\textbf{quality factor} \\
\textbf{oscillator~No.~1}& $\approx$ 20 cm$^{-1}$ & $\approx$ 3.0 cm$^{-1}$ & $\approx$ 0.93 \\
\textbf{oscillator~No.~2}& $\approx$ 37 cm$^{-1}$ & $\approx$ 5.9 cm$^{-1}$ & $\approx$ 1.35 \\
\textbf{oscillator~No.~3}& $\approx$ 72 cm$^{-1}$ & $\approx$ 9.7 cm$^{-1}$ & $\approx$ 1.89 \\
\textbf{oscillator~No.~4}& $\approx$ 118 cm$^{-1}$ & $\approx$ 7.8 cm$^{-1}$ & $\approx$ 4.00 \\
\textbf{oscillator~No.~5}& $\approx$ 142 cm$^{-1}$ & $\approx$ 2.8 cm$^{-1}$ & $\approx$ 9.00 \\
\textbf{oscillator~No.~6}& $\approx$ 190 cm$^{-1}$ & $\approx$ 16.5 cm$^{-1}$ & $\approx$ 5.00 \\
\textbf{oscillator~No.~7}& $\approx$ 237 cm$^{-1}$ & $\approx$ 10.4 cm$^{-1}$ & $\approx$ 8.80 \\
\textbf{oscillator~No.~8}& $\approx$ 260 cm$^{-1}$ & $\approx$ 6.1 cm$^{-1}$ & $\approx$ 10.80 \\
\textbf{oscillator~No.~9}& $\approx$ 282 cm$^{-1}$ & $\approx$ 9.9 cm$^{-1}$ & $\approx$ 11.75 \\
\textbf{oscillator~No.~10}& $\approx$ 325 cm$^{-1}$ & $\approx$ 4.8 cm$^{-1}$ & $\approx$ 18.06 \\
\textbf{oscillator~No.~11}& $\approx$ 363 cm$^{-1}$ & $\approx$ 6.3 cm$^{-1}$ & $\approx$ 20.17 \\
\textbf{oscillator~No.~12}& $\approx$ 380 cm$^{-1}$ & $\approx$ 5.3 cm$^{-1}$ & $\approx$ 29.23 \\
\textbf{oscillator~No.~13}& $\approx$ 426 cm$^{-1}$ & $\approx$ 4.4 cm$^{-1}$ & $\approx$ 30.43 \\
\textbf{oscillator~No.~14}& $\approx$ 478 cm$^{-1}$ & $\approx$ 3.4 cm$^{-1}$ & $\approx$ 48.00 \\
\textbf{oscillator~No.~15}& $\approx$ 500 cm$^{-1}$ & $\approx$ 1.3 cm$^{-1}$ & $\approx$ 35.71 \\
\hline
 \\
&\multicolumn1c{\hspace{3cm} \bf{Quantum simulator}}\\
\noalign{\vskip-11pt}
\\
\hline 
&\textbf{transition frequency} &\textbf{coupling strength}&\textbf{quality factor} \\
\textbf{oscillator~No.~1}& $\approx$ 120 MHz & $\approx$ 18.00 MHz & $\approx$ 0.93 \\
\textbf{oscillator~No.~2}& $\approx$ 222 MHz & $\approx$ 35.38 MHz & $\approx$ 1.35 \\
\textbf{oscillator~No.~3}& $\approx$ 432 MHz & $\approx$ 58.16 MHz & $\approx$ 1.89 \\
\textbf{oscillator~No.~4}& $\approx$ 707 MHz & $\approx$ 46.77 MHz & $\approx$ 4.00 \\
\textbf{oscillator~No.~5}& $\approx$ 851 MHz & $\approx$  16.79 MHz & $\approx$ 9.00 \\
\textbf{oscillator~No.~6}& $\approx$ 1.1 GHz & $\approx$ 98.93 MHz & $\approx$ 5.00 \\
\textbf{oscillator~No.~7}& $\approx$ 1.4 GHz & $\approx$ 62.36 MHz & $\approx$ 8.80 \\
\textbf{oscillator~No.~8}& $\approx$ 1.6 GHz & $\approx$ 36.57 MHz & $\approx$ 10.80 \\
\textbf{oscillator~No.~9}& $\approx$ 1.7 GHz & $\approx$ 59.36 MHz & $\approx$ 11.75 \\
\textbf{oscillator~No.~10}& $\approx$ 1.9 GHz & $\approx$ 28.78 MHz & $\approx$ 18.06 \\
\textbf{oscillator~No.~11}& $\approx$ 2.2 GHz & $\approx$ 37.77 MHz & $\approx$ 20.17 \\
\textbf{oscillator~No.~12}& $\approx$ 2.3 GHz & $\approx$ 31.79 MHz & $\approx$ 29.23 \\
\textbf{oscillator~No.~13}& $\approx$ 2.6 GHz & $\approx$ 26.38 MHz & $\approx$ 30.43 \\
\textbf{oscillator~No.~14}& $\approx$ 2.9 GHz & $\approx$ 20.39 MHz & $\approx$ 48.00 \\
\textbf{oscillator~No.~15}& $\approx$ 3 GHz & $\approx$ 7.79 MHz & $\approx$ 35.71\\
\hline
\end{tabular}
\caption{\label{Exp-Para23}Decomposition of the temperature-dependent experimental
\cite{wendling2000} mode density at 300~K shown in Figure~4~(b) of the main article and simulation 
with 15 LRC-oscillators coupled to each flux qubit, see Figure~2~(a) of the main article.}
\end{table*}

%%%%%%%%%%%%%%%%%%%%%%%%%%%%%%%%%%%%%%%%%%%%%%%%%%%%%%%%%%%%%%%%%%%%%%%%%%